\documentclass[conference]{IEEEtran}
\IEEEoverridecommandlockouts
\usepackage{cite}
\usepackage{amsmath,amssymb,amsfonts}
\usepackage[super]{nth}
\usepackage{url}
\usepackage{algorithmic}
\usepackage{graphicx}
\usepackage{textcomp}
\usepackage{xcolor}
\usepackage{array}
\def\BibTeX{{\rm B\kern-.05em{\sc i\kern-.025em b}\kern-.08em
    T\kern-.1667em\lower.7ex\hbox{E}\kern-.125emX}}
\begin{document} 
\pagenumbering{roman}



\title{Correct and Control Complex IoT Systems: Evaluation of a Classification for System Anomalies\\}


\author{\IEEEauthorblockN{1\textsuperscript{st} Sina Niedermaier}
\IEEEauthorblockA{\textit{Institute of Software Technology} \\
\textit{University of Stuttgart}\\
Stuttgart, Germany \\
sina.niedermaier@iste.uni-stuttgart.de}
\and
\IEEEauthorblockN{2\textsuperscript{nd} Stefan Heisse}
\IEEEauthorblockA{\textit{Bosch Engineering GmbH,} \\
\textit
Stuttgart, Germany\\
stefan.heisse@de.bosch.com}
\and
\IEEEauthorblockN{3\textsuperscript{rd} Stefan Wagner}
\IEEEauthorblockA{\textit{Institute of Software Technology} \\
\textit{University of Stuttgart}\\
Stuttgart, Germany \\
stefan.wagner@iste.uni-stuttgart.de}
}

\maketitle

\begin{abstract}
In practice there are deficiencies in precise inter-team communications about system anomalies to
perform troubleshooting and postmortem analysis along different
teams operating complex IoT systems. 
We evaluate the quality in use of an adaptation of IEEE Std. 1044-2009 with the
objective to differentiate the handling of fault detection and fault reaction
from handling of defect and its options for defect correction. We extended the
scope of IEEE Std. 1044-2009 from anomalies related to software only to
anomalies related to complex IoT systems.
To evaluate the quality in use of our classification a study was conducted at
Robert Bosch GmbH. We applied our adaptation to a postmortem analysis of an IoT
solution and evaluated the quality in use by conducting
interviews with three stakeholders. Our
adaptation was effectively applied and inter-team
communications as well as iterative and inductive learning for product improvement were enhanced. Further training and practice are required.

\end{abstract}


\section{Introduction}
\label{sec:introduction}

Providers of cloud-based IoT solutions compile distributed complex systems of components, integrating hardware, software and mechanical system elements. In operations they aim for controlling their dynamic behaviour in a permanently changing context of use.
The system components are developed and operated by different teams from different organizational units. The teams are often specialized in the context of their component and their organizational silo. Communication within an organizational silo and the teams is performed by applying concepts that are optimized for their specific intra-team context.

In case of failure, a failed function of an IoT service,
the different teams operating the system components, have to form a collective to collaboratively control the situation. 

\textbf{First}, in situation management (troubleshooting), facing dynamics, the objective is to \textbf{regain control} as quickly as possible \textbf {with fault reactions}. The purpose of fault reaction is to avoid a failure influencing customer satisfaction or, if not successful, at least to provide limited operation for keeping the customer's trust by controlling the situation. 

\textbf{Second}, once the system is transitioned back into a controlled state, the anomaly in the effect chain and the actions of responding to dynamic cascades of consecutive anomalies are investigated within a postmortem analysis. The objective is to \textbf{detect defects} and subsequently \textbf{remove them}. \\

Each of the two sets of activities, situation management and postmortem analysis follow different optimization goals.

\textbf{First}, detection of service failure initiates \textbf{tactical} actions performed 
with the goal of customer satisfaction. The leading principle is to 'be \textbf{effective} before efficient'.

\textbf{Second}, the subsequent analysis to identify defects and options for correction initiates \textbf{strategical} actions 
performed with the goal of profitability. The leading principle is to 'be \textbf{efficient}'.\\

To perform situation management and postmortem analysis 
several development and operational teams from different organizational silos form a collective\footnote{Set of diverse participants to persist in complex domains.}. Deficiencies on shared concepts for communication about anomalies often lead to misunderstandings and different interpretations regarding anomalies.
Missing shared concepts on anomalies for inter-team communications across organizational silos can be used to direct ``responsibility"  for failure, its compensation and related defect correction to a specific organizational unit \cite{niedermaier2019observability}.

Assigning blame to a specific organizational unit can be intensified by performing root cause analysis, for example by asking 5xWhy \cite{ohno1988toyota}. While doing this, people tend to search for isolated individual causes. In retrospective, humans strive to oversimplify causal chains to a single root cause without embracing complexity and dynamics in cause effect chains \cite{kahneman2011thinking}.

As stated by Cook \cite{cook1998complex} for distributed complex systems there might be no isolated root cause. 
Rather, there are multiple faults and contributors, where each fault is necessary but they are only jointly sufficient for failure \cite{cook1998complex}. 
Asking ``why" tends to point in the direction of ``who". This oversimplification to a single root cause and a single responsibility often results in a culture of blaming.

In contrast, a retrospective is an instrument of a learning organization performing feedback cycles to improve product and service quality as well as organization quality\cite{SengeLearning, forrester1994system, DINENISO9000}. The feedback cycles process defects related to product and service quality and defects related to the performance of the~organization. \\
If the differentiation of effective tactical work mode from efficient strategic work mode is missing, an organization tends to remain in effective tactical mode. If a failure can be controlled only by tactical \textbf{fault} reactions, an interest in the defect and its sustainable correction often remains secondary. Therefore, \textbf{defects} with their options for correction are not listed and accumulate to technical debts.




To address this need for a precise differentiation 
and to enhance the effectiveness and efficiency of the collaboration capabilities of developers, operators and organizations, we propose a shared classification for anomalies.
This work is based on the common logical concept of the IEEE Std. 1044-2009 \cite{ClassificationSoftwareAnomalies} to classify software anomalies. 

We propose an adaptation of the IEEE Std. 1044-2009 applied only to system element software to extend the same for application to a system composed of a set of interacting system elements. This enables differentiation between the handling of fault detection and dynamic fault reactions from the handling of defect and its options for sustainable defect correction, not only on system element software but generally on system level.

We investigated its quality in use \cite{ISOIEC25022} by conducting a case study with two stages. In the first stage we applied our adaptation of the IEEE Std. 1044-2009 to a postmortem analysis and present the results of the classification. To further evaluate the quality in use of our adaptation, we conducted semi-structured interviews with three of the stakeholders of the postmortem analysis.\\

The remaining part of this work is organised as follows: The objective of this work -- the evaluation of the quality in use of our adaptation of the IEEE Std. 1044-2009 to differentiate between the handling of fault and fault reaction from the handling of defect and its options for defect correction -- are presented in Section~\ref{research objective}. We further explain the methodology of the two-staged case approach that we use to evaluate our adaptation. In Section~\ref{related work} we give an overview of related work in the areas of classification for anomalies and postmortem analysis. The concepts of this work are presented in Section~\ref{Concept} and include a system definition with three architectural views: functional system view, dynamic system view and system structure view. In addition, we introduce the concepts used in IEEE Std. 1044-2009: problem, failure, fault and defect. Based on our three system views, we propose our adaptation of the concepts fault and defect and further introduce the concept of fault tolerance and situation management as well as defect correction and quality management. In Section~\ref{section_case}, we illustrate the results of our case study on the quality in use of our adaptation and conclude the paper with a discussion of implications as well as limitations and propose future work.
\\

\section{{Scope and Research Method}}
To structure our research, we applied the case study research process as proposed by Runeson and Hoest \cite{runeson2009guidelines}. With the case study we addressed the following research goal.

\subsection{Research Goal} \label{research objective}
The goal of this study is to evaluate the quality in use (with limitation to the quality sub-characteristics: effectiveness, efficiency and satisfaction) of an adaptation of the Standard Classification for Software Anomalies, the IEEE 1044-2009, by applying it to a postmortem analysis of an IoT System. 

As part of our adaptation, we have extended the scope of IEEE Std. 1044-2009 from  anomalies related to software only to anomalies related to complex IoT systems providing service to a customer. 

We differentiate between the handling of: 
\begin{itemize}
\item {first, \textbf{fault }detection and \textbf{fault reaction}}

in the context of \textbf{situation management} from
\end{itemize}
 
 \begin{itemize}
 \item {second, \textbf{defect} and its options for \textbf{defect correction} }
 
 in the context of \textbf{defect management}.\\
 \end{itemize}

This work is intended to enhance effectiveness and efficiency of the collaboration capabilities between developers, operators along organizational units by providing common logical concepts of anomalies. The concepts of anomalies enable precise inter-team communication to differentiate between actions for controlling the system and actions for its sustainable correction.\\

For evaluation of the quality in use with its quality sub-characteristics effectiveness, efficiency and satisfaction the following definitions are presented and interpreted for the context of this work.

According to ISO/IEC 25022 \cite{ISOIEC25022} \textbf{quality in use} is the: 
``degree to which a product or system can be used by specific users to meet their needs to achieve specific goals with effectiveness, efficiency, satisfaction, and freedom from risk in specific contexts of use". For the context of this work, we focus on the sub-characteristics of effectiveness, efficiency and satisfaction as defined in ISO/IEC 25022 \cite{ISOIEC25022}.
\begin{itemize}
\item \textbf{Effectiveness:} ``Accuracy and completeness with which users achieve specified goals" \cite{iso-9241-11-usability}.
We evaluate the accuracy of differentiating the handling of fault detection and fault reaction from the handling of defect detection and its defect correction option(s), according to our adapted classification. The evaluation of the completeness in distinguishing the anomalies is not necessary due to an unlimited amount of possible defects as described in Section~\ref{discussion- stage1}.

\item \textbf{Efficiency}: ``Resources expended in relation to the accuracy and completeness with which users achieve goals"~\cite{iso-9241-11-usability}. Resources include temporal and also mental effort performing the task.
Since we performed qualitative research we did not track time consumption measurements. We therefore qualitatively evaluated the estimated efficiency of the interview participants in comparison to other postmortem analyses.

\item \textbf{Satisfaction}: ``Degree to which user needs are satisfied when a product or system is used in a specified context of use" \cite{ISOIEC25022}.
 We evaluated this characteristic by asking the interview participants about their subjective opinions and attitude towards applying the classification.\\ 
\end{itemize}

\subsection{Case Context}
We evaluate the quality in use of the classification for system anomalies in a case study conducted in 2019 at Robert Bosch GmbH, a German company providing IoT solutions. 

The case study is based on a postmortem analysis related to a customer problem in the domain of IoT condition monitoring. We designed the case study in a two-staged procedure. In stage 1, we applied the adapted classification on a real-world postmortem analysis of the IoT solution. In stage 2, we performed semi-structured interviews to evaluate the quality in use of our concept with the stakeholders of the postmortem analysis.

\subsection{Stage 1 Application - Data Collection and Analysis}
The data collection and analysis of stage one was conducted between February and May 2019 in the case company. 
In this stage specialists, representing the stakeholders interest formed a collective to perform postmortem analysis. For analysis, the method of 5xWhy was executed to assign root causes.
Selected stakeholders of the collective applied our classification for anomalies to enable intra-collective communications.

The data collection technique can be determined as first degree where two of the researchers were in direct relation to the case company. One researcher acted as coach to ensure the application of our classification during the postmortem analysis. The second researcher took the role of an independent observer. For data triangulation reasons, we included data from multiple sources by collecting incident report data, notes from postmortem analysis meetings, observations and mails discussing the classification for anomalies. 
As the data is sensitive it is not provided with this work.\\ 
We analysed the data by performing data triangulation~\cite{stake1995art}. By member checking, the results were revised by the stakeholders of the postmortem analysis and have been simplified for the purpose of this article in Section~\ref{results_classification}.\\


\subsection{Stage 2 Quality in Use Interviews - Data Collection and Analysis}
To further evaluate the quality in use of the proposed classification for anomalies, we performed semi-structured interviews with one participant of each stakeholder group of the postmortem analysis. The interviews aim to explore the individual experiences of the interviewees and to qualitatively evaluate the quality in use of our classification. All of the interviewees were actively involved during the postmortem analysis and are responsible for the resulting classification. The stakeholders include the product manager (S1) of the IoT solution, a quality manager of the cloud infrastructure provider (S2) and a systems analyst and quality expert (S3), who has coached the other stakeholders in applying the classification (see Table \ref{tab:Stakeholder_Demographics}). All of the three interviewees are employees of the company. \\

\begin{table}[ht!]
\scriptsize

    \caption{Interviewee Information}
    \label{tab:Stakeholder_Demographics}
   \begin{tabular}{m{0.5cm} m{3.5cm} m{3.5cm}} 
      \hline
     \textbf{SID} &  \textbf{Role} &  \textbf{Focus in the Analysis} \\
      \hline
     S1 & Product Manager of IoT Solution & Representing customer perspective \\
     S2 & Quality Manager of Cloud Provider & Lessons learned \\
     S3 & Quality Expert Consultant & Systems analysis\\
      \hline
      \multicolumn{3}{c}{*SID = Stakeholder ID}
    \end{tabular}
 
\end{table}

For semi-structuring the interviews, we created an interview guide \cite{niedermaier2019guideline}. The interview guide is structured as follows:
\begin{itemize}
    \item general questions about classifications for anomalies 
    \item main blocks about the effectiveness, efficiency and satisfaction of the application of our classification for anomalies
    \item open questions to discover the pros and cons of our classification
    \item open question to identify potential for optimization.
\end{itemize}
We loosely followed the questions and audio recorded the three face-to-face interviews of approximately 40 minutes, which were all held in German. After transcribing, we sent the interviews to the participants for review.\\
For evaluating the quality in use of the classification, we analysed the semi-structured interviews by performing Mayring's approach of qualitative content analysis \cite{mayring2014qualitative}. We performed a mixed approach of deductive and inductive coding by encoding the transcripts with the predefined quality in use criteria of Section~\ref{research objective} and creating further categories for contents which could not be directly assigned to the existing categories. The transcripts were analyzed on a sentence level. During analysis, we formed hierarchies of codes and sub-codes. Through several iterations, the codes were revised, spilt or merged. The results of the interviews are described in Section~\ref{results_interview}.\\

\section{\uppercase{Related Work}\label{related work}}
To provide context to this study, the related work investigates classifications for anomalies, as well as the method of postmortem for analysing service disruptions and investigating the handling of anomalies. 

\subsection{Classifications for anomalies}\label{relatedWorkClassification}
According to Wagner \cite{Wagner:2008:DCD:1390817.1390829} there are several anomaly and defect classifications with different purposes, e.g. to enhance the identification of defects or the education of developers. One of them is even an IEEE standard, the IEEE Std. 1044-2009~\cite{ClassificationSoftwareAnomalies} classifying software anomalies. The standard introduces the different concepts of problem, failure, fault and defect. The IEEE Std. 1044-2009 is the basis for our classification, therefore we introduce it in detail in Section~\ref{section classification of anomalies}.

An application of the IEEE Std. 1044-2009 can be found in the case study of Mellegard et al. \cite{mellegard_staron_volvo}, within the context of automotive software development. The authors focused on the recognition and analysis of the defects. In contrast to their work, we aim to differentiate between the handling of fault detection and fault reaction from the handling of defect and its options for defect correction.

Another widespread classification for anomalies is the taxonomy of dependable and secure computing by Avizienis et al. \cite{avizienis2004basic}. The authors address the threats to dependability and security by the concepts of failure, error and fault. Additionally, Avizienis et al. apply fault tolerance techniques by handling errors and faults. 
The classification of this work has several symmetries to the classification of Avizienis et al. It is possible to roughly map their concepts of failure, error and fault to the concepts of failure, fault and defect of this work. The main difference of the concept of defect of this work to their concept of fault is that they consider the fault in a static system view as well as in a dynamic system view and do not differentiate between them. 

\subsection{Postmortems}\label{relatedWorkPostmortem}
With  the concept of postmortems, critical service disruptions are analyzed in retrospective with a focus on how well the organization responded and recovered from the disruption. Postmortems are an essential part of Google's discipline of Site Reliability Engineering \cite{beyer2016site}, that is increasingly being used in various companies \cite{niedermaier2019observability}.
Part of the approach of postmortems is the concept of root cause analysis (RCA) \cite{beyer2016site}. RCA aims to identify the root of a an anomaly and also why it was introduced. Thereby, RCA targets to prevent similar causes from being introduced in the future \cite{Wagner:2008:DCD:1390817.1390829}. As we refer to the work of Cook \cite{cook1998complex}, who asserts that there is no root cause in complex system, with this work, we try to avoid the concept of searching for single causing entity, but rather to identify defects with their correction options. 
\\


\section{\uppercase{Concepts}} \label{Concept}
In this section, we introduce the concepts of anomalies, applied in our classification. We provide a system definition and present the concepts of the IEEE Standard Classification for Software Anomalies \cite{ClassificationSoftwareAnomalies} and our adaptation for complex systems providing IoT services. 
Furthermore, we introduce the concept of fault tolerance and situation management, where we focus on the relationship between fault and failure to keep the system in a controlled state. We complete this section by presenting the concept of defect correction and quality management, where we focus on the relationship between fault and defect for sustainable defect correction, optimized for strategic interest.

\subsection{Classification for Anomalies} \label{section classification of anomalies}
The IEEE Std. 1044-2009 is a standard which provides a unified approach to classify software anomalies and models the relation between software anomalies and maintenance activities \cite{ClassificationSoftwareAnomalies}. Our adaptation of the concepts of the IEEE Std. 1044-2009 is general enough to cover not only \textit{software} but also anomalies related to \textit{complex IoT systems}. A \textbf{system} in this work is defined by its components, compiled into a structural arrangement interacting in component-effect relationships. The components itself are composed of interacting system elements such as software, hardware, mechanics, humans, procedures, etc. \cite{ISOIEC24748}.\\

We present the concepts of software anomalies of IEEE Std. 1044-2009 \cite{ClassificationSoftwareAnomalies} and differentiate our concepts applicable to complex systems providing IoT services.\\

\begin{itemize}
\item \textbf{Problem}: Difficulty or uncertainty experienced by one or more persons, resulting from an unsatisfactory encounter with a system in use.
\item \textbf{Failure}: An event in which a system or system component does not perform a required function within specified limits. (adapted from ISO/IEC 24765:2009 \cite{ISOIEC24765})
\item \textbf{Fault}: A manifestation of an error in software. (adapted from ISO/IEC 24765:2009 \cite{ISOIEC24765})
\item \textbf{Defect}: An imperfection or deficiency in a work product where that work product does not meet its requirements or specifications and needs to be either repaired or replaced. (adapted from the Project Management Institute \cite{PMI})\\
\end{itemize}

\begin{figure}[ht!]
	\centering
	\includegraphics[width=.48\textwidth]{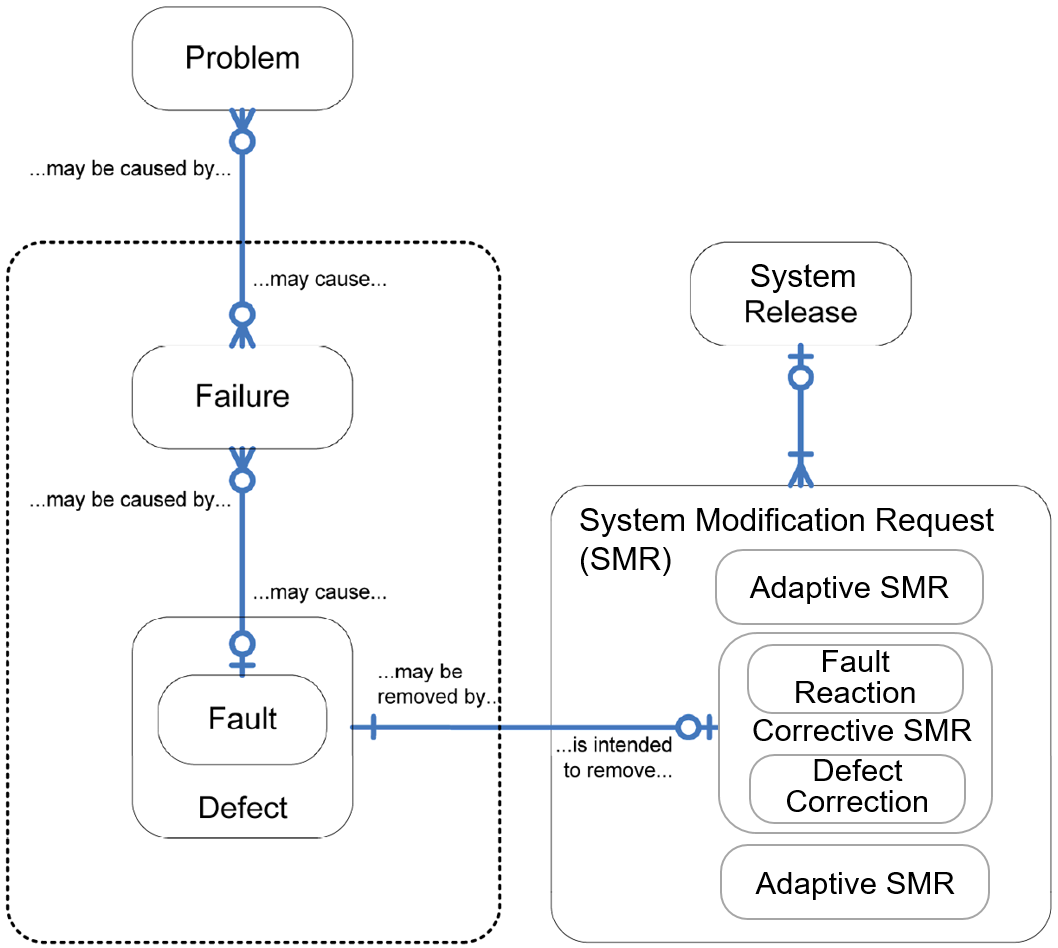}
	\caption{Relationships modeled as an entity relationship diagram (adapted from~\cite{ClassificationSoftwareAnomalies})}
	\label{img:figure_relationship_problem_failure_fault_defect}
\end{figure}

The relationships between the different concepts are shown in Figure~\ref{img:figure_relationship_problem_failure_fault_defect}. 
Further the IEEE Std. 1044-2009 ~\cite{ClassificationSoftwareAnomalies} depicts the relationship between the concepts \textbf{fault} and \textbf{defect} as follows:\\
\textit{``A fault is a subtype of the supertype defect.
Every fault is a defect, but not every defect is a fault.
A defect is a fault if it is encountered during software execution (thus causing a failure). A defect is not a fault if it is detected by inspection or static analysis and removed prior to executing the software."}\\

For developing and operating complex IoT systems three architectural views are motivated. ''A view is a representation of a whole system from the perspective of a related set of concerns" \cite{ISOIEC42010} .

\begin{itemize}
\item \textbf{System structure view:} The system structure view presents the components in their structural arrangement. The system structure view is optimized to organize the division of labour.   
\item \textbf{Functional system view:} The functional view breaks down the customer function into sub-functions that can be assigned to the system components. The functional system view is optimized to represent the value proposition to the customer. 

\item \textbf{Dynamic system view:} The dynamic view represents the component effect\footnote{For this work, component effect relationship is equivalent to cause effect relationship. A cause is related to a component, therefore we use the concept of component effect chain.} relationships in temporal expanse. The dynamic system view is optimized for situation management to control the dynamic behaviour. 
\\
\end{itemize}

To avoid conflicts with other concepts of anomalies, we choose a deductive approach and motivate the concepts of anomaly out of the system views.
The views contain conformities and nonconformities. 
We define a nonconformity related to a system view as an anomaly.
The anomalies of IEEE Std. 1044-2009 can be mapped to the anomalies of the three system views.


\begin{itemize}
\item \textbf{Failure}: A failure is a failed function, where a  system  does  not  perform  a  required  function within specified limits. This concept of anomaly is assigned to and optimized for the \textbf{functional system view}. 
\item \textbf{Fault}: A fault is a break in a component effect chain and has a temporal expanse. This concept of anomaly is assigned to and optimized for the \textbf{dynamic system view}. A fault can be classified as permanent, transient or intermittent \cite{ISO26262FunctionalSafety}.
\item \textbf{Defect:} Whereas the defect in IEEE Std. 1044-2009 is an imperfection or deficiency in a work product, we assign the defect to a subset of components in their structural arrangement and identify \textit{potential correction option(s)}. Option(s) for correction are related to modification on a:
\begin{itemize}
    \item subset of components
    \item component itself
    \item system element of a component
\end{itemize}
The trivial correction option is the exchange of the subset of components the defect is assigned to. An assignment of correction option(s) to a subset of components does not imply poor quality of work.
The concept of anomaly defect and its correction option(s) are assigned to and optimized for the \textbf{system structure view.}\\
\end{itemize}

We adapted the software change request (IEEE 1044-2009) to a system modification request (SMR)\footnote{For this work, the system modification request includes the software change request. A SMR in virtual systems focuses on a re-compiled (corrected) structural re-arrangement of a subset of components and not on modifying system elements of components.}\cite{IEE14764-2006} assigned to a subset of components in structural arrangement. 

As a fault is a defect, there is a (see Figure \ref{img:figure_relationship_problem_failure_fault_defect}):
\begin{itemize}
    
\item \textbf{\nth{1}} corrective SMR resulting in a (quick) \textbf{fault fix} and a

\item \textbf{\nth{2}} corrective SMR resulting in a \textbf{defect correction}. 
\end{itemize}


\subsection{Fault Tolerance and Situation Management} \label{section fault tolerant time intervals}
Within this section, we focus on the relationship between the concept fault and concept failure. 
In engineering complex IoT systems we aim for fault tolerance: ``the ability to deliver a specified functionality in the presence of one or more specified faults" \cite{ISO26262FunctionalSafety}. 
Ideally, a system is designed fault tolerant. If the system is not tolerant to a fault, we face a situation: we have to react to keep customer's trust.
Now time matters. We are bound to time and switch into effective mode. \\

\begin{figure}[ht!]
	\centering
	\includegraphics[width=.5\textwidth]{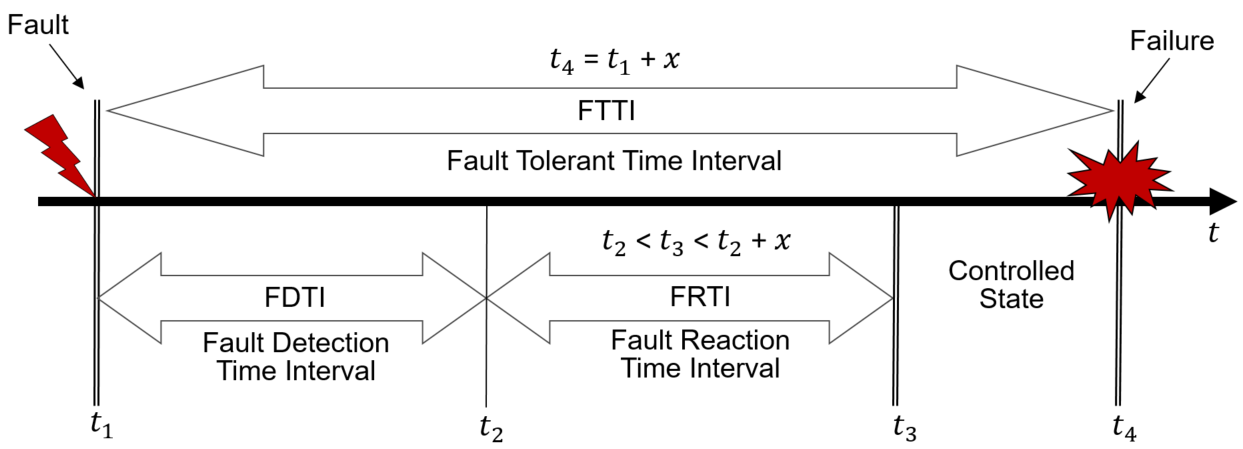}
	\caption{Fault tolerant time interval (FTTI) (adapted from~\cite{ISO26262FunctionalSafety})}
	\label{img:fault_tolerant_time_interval}
\end{figure}

Analogous to ISO 26262 -- Functional safety -- applied in engineering control systems \cite{ISO26262FunctionalSafety}, we differentiate the following time stamps and time intervals in the \textbf{dynamic system view} (see Figure \ref{img:fault_tolerant_time_interval}).

Time stamps:
\begin{itemize}
\item\(t_1\): Occurrence of fault, fault is not detected.
\item\(t_2\): Time when fault is detected.
\item\(t_3\): End of fault reaction. 
\item \(t_4\): End of fault tolerant time interval. Occurrence of failure.
\end{itemize}

Time intervals:
\begin{itemize}
\item \textbf{FTTI} (Fault tolerant time interval) is the minimum time span from the occurrence of a fault in a component-effect chain of a system (dynamic view) to a possible occurrence of a failure if the quality mechanism is not activated.
\item \textbf{FDTI} (Fault detection time interval) is the time span from the occurrence of a fault to its detection.
\item \textbf{FRTI} (Fault reaction time interval) is the time span from the detection of a fault executing a fault reaction until a transition to a controlled state is achieved.
\end{itemize}

\textbf{Fault Tolerance:} For improving fault tolerance we automate fault reactions. Fault reaction is performed to prevent the fault from progressing to failure.
The fault is detected in FDTI. Fault reaction is performed in FRTI and has to be successful before FTTI ends.

\textbf{Situation management:} If the system is not tolerant to a fault, detecting the fault we switch into a tactical mode by performing an OODA loop \cite{boyd1996essence}. We \textbf{O}bserve, \textbf{O}rient, \textbf{D}ecide and \textbf{A}ct (OODA) in the dynamic system view. 
For keeping the customer's trust we demonstrate that we control the situation.
In FTTI, we transfer the system into a controlled state that is predefined and agreed for customer context of use.
We engineer in having tactical options: we prepare quality mechanisms, fault reactions which transfer the system into a controlled state.
Controlled states include:
\begin{itemize}
\item  \textbf{redundancy activated:} no failure in customer context~\cite{avizienis2004basic}.
\item \textbf{degraded}: permission to use a service that does not conform to specified requirements \cite{DINENISO9000} e.g. graceful degradation by serving stale data\cite{dekker2008resilience}. 
\item \textbf{regraded:} alteration of the grade of a nonconforming service in order to make it conform to requirements differing from the initial requirements (related to value proposition) \cite{DINENISO9000}.
\item \textbf{terminated:} preclude usage by controlled take out of service \cite{DINENISO9000} e.g. isolation. \cite{avizienis2004basic}
\item \textbf{compensated:} compensation of damages for the customer, e.g. in the form of monetary award.
\end{itemize}

In situation management the customer is watching us and assesses behaviour and communications and decides to continue placing trust on us. 
Collaborating with the customer, precise communication about anomalies relies on shared concepts.\\

If the (quick) fault fix is effective, the system is in a controlled state. We control the situation. The customer trust remains placed on us. 
Now we are independent from time, we leave the dynamic system view. We switch into efficient mode. We perform quality management, in particular defect management in a PDCA loop (\textbf{P}lan, \textbf{D}o, \textbf{C}heck, \textbf{A}ct) \cite{DINENISO9000}.
During a postmortem analysis, improvement for fault reaction to increase fault tolerance as well as improvement for tactical options to increase capability for situation management can be identified and transferred to defect management.


\subsection{Defect Correction and Quality Management (Defect Management)}
Sustainable defect correction, which is optimized for strategic interest is not equivalent to a (quick) fault fix, which is optimized for tactical interest.
Now, we act in efficient mode with the goal of profitability. \\

\begin{figure}[ht!]
	\centering
	\includegraphics[width=.48\textwidth]{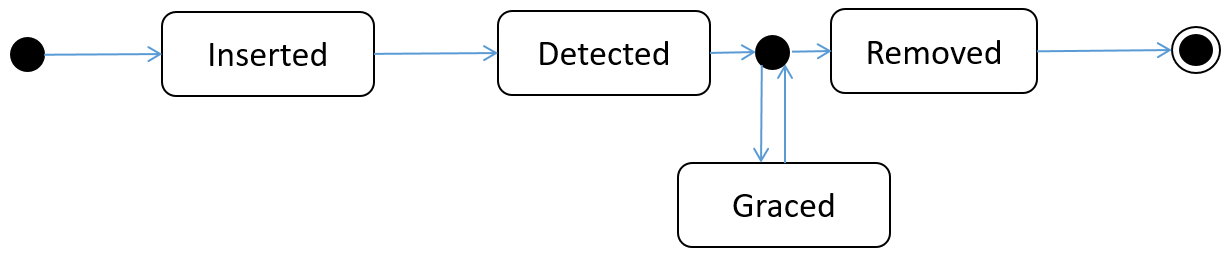}
	\caption{Defect life cycle (adapted from~\cite{ClassificationSoftwareAnomalies})}
	\label{img:figuredefect_lifecycle_adapted}
\end{figure}

The lifecycle of a defect is depicted in Figure~\ref{img:figuredefect_lifecycle_adapted}.

In postmortem analysis, we identify anomalies related to the different system views:
\begin{itemize}
\item In the \textbf{functional system view} an anomaly \textbf{(failure)} with need for correction is identified, a \textbf{defect is detected} and is to be assigned to system structure view. 
\item In the \textbf{dynamic system view }an anomaly \textbf{(fault)} with the need for correction is identified, a \textbf{defect is detected} and is to be assigned to system structure view.
\item In the \textbf{system structure view} an anomaly\textbf{ (defect)} with the need for correction is detected and is to be assigned to a subset of components.  
\end{itemize}{}
 

As described in Section \ref{section classification of anomalies}, a defect and its option(s) for correction are assigned to \textbf{system structure view} and a subset of components in their structural arrangement.
The defect management decides whether and which defect correction option, addressed by a SMR, is to be executed for strategic and efficiency interest. \\



In Figure~\ref{img:figuredefect_lifecycle_adapted}, we adpated the UML statechart diagram of the defect life cycle of the IEEE Std. 1044-2009~\cite{ClassificationSoftwareAnomalies}.
The different states of a defect \textit{inserted, detected, removed} have been extended with an additional state \textit{graced}. We propose, that after a fault is detected, there is the additional option to grace a defect for a certain period of time until it is removed. 
The decision to grace a defect is based on an assessment that a defect of an complex IoT system is accepted under following circumstances:
\begin{itemize}
\item The decision can take place due to economical reasons, when the organization decides to remain in a fault reaction mode (as described in Section~\ref{section fault tolerant time intervals}) instead of sustainably remove the defect. 
\item The decision to grace a defect can be influenced  by the assumption that there will be acceptable customer impact.
\item The decision to grace a certain defect may be necessary in the context of a safety vs. security vs. privacy discussion.
For example, a defect from security perspective -- a security defect -- may be graced for safety or privacy reasons. 
\end{itemize}

Nevertheless, defects have to be documented in a defect list and are to be transferred to defect management. 

\subsection{Communicating about Anomalies}

The relationship between the different concepts of Section~\ref{section classification of anomalies} can be used to communicate about the anomalies along different organizational units as follows:\\

\textit{The customer got a \textbf{Problem}
caused by a  \textbf{failure} (a failed function which is required, which is promised with a value proposition), 
indicating the presence of a permanent, intermittent or transient \textbf{fault} (activated by customer context of use). 
The fault has to be detected and a fault reaction has to be executed to transfer the system into a controlled state.
The fault is indicating the presence of a \textbf{defect}, which has to be assigned to a subset of components and is intended to be removed by a completed defect correction (of a corrective SMR).}\\

\section{\uppercase{Case Study}} \label{section_case}
We evaluate quality in use of the concepts by applying them to the postmortem analysis of the IoT solution. This section provides an abstract description of the IoT solution and the context of the customer problem. Further, we present the results of the classification (stage 1) and the results of the interview analysis (stage 2).

\subsection{IoT System Description}\label{system overview}
The mission of the IoT solution can be stated as follows:\\
\textit{Generate and provide condition monitoring data of a physical customer asset to increase efficiency in managing it.} 

The IoT solution relies on a distributed system architecture, with relation to different stakeholders: customer, solution provider, cloud infrastructure provider, sensor gateway provider and sensor provider.

\subsection{Situation}\label{problem description}
During operations the solution provider was not able to perceive faults in the effect chains and the instantiation of a related failure of the complex IoT system.
The customer notification of failure was the only indicator of faults.

The customer informed the organization, calling the service desk, that a required function had failed.
The incident-specific problem solving procedure was not initiated before customer notification. The problem solving procedure ran by the company, includes the initiation of fault reactions to transfer the system into a controlled state. A team of specialists, representing the stakeholders interest, was executing postmortem analysis, applying the method of 5xWhy to assign root causes.
In the analysis the stakeholders applied our classification for anomalies to communications.


\subsection{Classification Results - Stage 1} \label{results_classification}
In the following, we describe the results of the case study. We applied the classification for anomalies to differentiate between the handling of:

\begin{itemize}
\item {first, \textbf{fault }detection and \textbf{fault reaction}}

in the context of \textbf{situation management} from
\end{itemize}
 
 \begin{itemize}
 \item {second, \textbf{defect} and its options for \textbf{defect correction} }
 
 in the context of \textbf{defect management}.\\
 \end{itemize}
 
\textbf{Problem}
\begin{itemize}
\item Customer view: customer does not receive up-to-date asset condition information. Loss of control, with risk of damage of customer asset.

\item Solution provider view: customer perceives that solution provider lacks competency to control the situation. The relationship of trust between customer and solution provider is at risk. \\
\end{itemize}

\textbf{Failure}
\begin{itemize}
\item Customer \textbf{(functional system view)}: failure of solution. Information related to asset condition inconsistent due to message transmission fault.

\item Provider\textbf{ (functional system view)}: failure of load balancer (single point of failure) with fail passive due to load test on cloud infrastructure. The load balancer failed by a temporary overload and reacted as expected by switching to its controlled  state\footnote{Load balancer is in it's controlled state -- no access -- which is optimized for governance but not for service delivery.}: The system is down. The server and its service is not available.\\
\end{itemize}

\textbf{Fault ($\rightarrow$ fault reaction)}
\begin{itemize}

\item Customer \textbf{(dynamic system view) }: -

Customer is only offered functional system view. Complexity of dynamic system view is hidden from the customer.
\item Provider\textbf{ (dynamic system view)}: failure and fault detection by customers via emergency call to service desk.
The organization did not detect intermittent transaction message faults.
In consequence, there was no fault reaction. Both, technical (redundancy) as well as an organizational (manual) fault reaction were missing.
Incident-based problem solving process did not start until customer calls or sends an email to the service desk.\\
\end{itemize}

\textbf{Defect ($\rightarrow$ defect correction option)}\\
Defects related to IoT system \textbf{(system structure view)}:
\begin{itemize}
\item Defect 1: malconfigured load balancer. 
\item Defect correction option 1: correction of load balancer malconfiguration.
\end{itemize}
Defects related to situation management capability \textbf{(system structure view)}:

\begin{itemize}
\item Defect 2: missing technical option for fault reaction of load balancer (single point of failure): no failing active and operational, by activating a redundant component.
\item Defect correction option 2: implement failing active and activating redundancy.
\item Defect 3: missing organizational options for fault reaction: no tactical options for situation management. On organizational level solution provider did not agree controlled states for degrading the value proposition. 
\item Defect correction option 3: implement degrading value proposition including customer information of quality degradation.
\item Defect ...
\end{itemize}

\subsection{Discussion - Stage 1}\label{discussion- stage1}
With the application of the classification for anomalies, we identified:
\begin{itemize}
\item by differentiating the concepts \textbf{fault} and \textbf{failure}, we improve in engineering and operating \textbf{fault tolerant} systems.
\item by differentiating the concepts \textbf{fault} and \textbf{defect}, 
after effective \textbf{fault fixing}, we are able to assign a defect to a component and identify its \textbf{defect correction} options for transfer to defect management. This is done 
in efficiency mode to run for efficiency yield.
We are customer oriented. 
Customer interest is attached to fault fixing and not to defect correction and decreasing technical dept. We identified that motivation for defect correction has to be provided by the organization.
\item by differentiating the \textbf{dynamic system view}, we are able to assign defects related to IoT system and defects related to capability of situation management. 
\item during postmortem analysis, we tended to assign the defect to a single component and not to an interacting subset of components. Therefore, we forgo correction options, with higher potential of efficiency yield and sustainability.
\item a need to increase observability in the dynamic system view with a focus on business process to control adding value to the customer.
\item for having appropriate tactical options in situation management, additional defects (related to situation management) can be assigned to a subset of components.\\
\end{itemize}






In this case the correction of load balancer malconfiguration (defect 1) covers defect correction and fault fix. Defect 2, to eliminate the single point of failure is transferred to backlog.

Improvement to
increase fault tolerance as well as improvement to increase capability for situation management have been identified and are transferred to defect management. 
The defect management decides whether and which correction options are to be executed.

\subsection{Interview Results - Stage 2} \label{results_interview}
This section provides the results of the participant interviewees where we further investigate the quality in use (as described in Section \ref{research objective}) applying our adaptation. 
Following the structure of the interview guide, we present the highest level of the coding system including purpose of classification for anomalies, the evaluation criteria (effectiveness, efficiency, satisfaction) and further optimization potential for our classification. We describe the different codes with exemplary statements from the stakeholders (S1, S2 and S3) of the postmortem analysis.\\

\textbf{Purpose of classification}:
At the beginning of each interview, we asked about the purpose of a classification for anomalies in the context of a postmortem analysis. All of the three participants described the purpose of it related to existing challenges. 

S1 and S2 outlined from experience with other postmortem analyses that discussions between different organizational units are often very imprecise if no shared concepts exist.
Further, S3 stated that: \textit{``In deciding conflicts of interest, it is advantageous to have the power of interpretation over the concepts of anomalies. This sometimes results in the rhetorically strongest person taking over the interpretation sovereignty over the concepts and, for example, directing the effort for corrections to other organizational units."} In terms of interpretation sovereignty a classification for anomalies allows to: \textit{``competently exchange about anomalies with colleagues from other organizational units} [...] \textit{and thus ensures that you have a common view on the situation and the system."} (S2). Further a commonly accepted classification, acts as \textit{``}[...] \textit{a common language} [...]", that enables to \textit{"}[...]\textit{{improve mutual understanding."}} (S2). For \textit{``}[...] \textit{yourself only you do not necessarily need a classification."} (S2).
 It is especially important in order to enable continuous learning for iterative and inductive product development and improvement along different organizations. \textit{"This is done via feedback loops. Learning for ``better" is done by excluding what you do not want to have. Learning for ``better" is done by excluding the anomalies. The challenging communication about anomalies across organizations is made easier by a common classification scheme."} (S3). S3 further explicates that 
 applying the concepts of anomalies enables the linking of a specific failure in the value proposition to the customer, into a direct causal relationship of components across silos and to jointly propose corrections.
S1 summarized the benefit of the adaptation of the classification as follows: \textit{``The differentiation is very important. When I detect a fault, I have to do a hot fix immediately. It's very important that I can get this to work at all. To correct it in a sustainable way, working on the defect probably makes sense out of a business perspective."}
To proceed with the defects from a business perspective, reusability of the results of the classification is necessary and was highlighted by all three stakeholders. \textit{"The information status for a decision, for dealing with the anomalies is available in a known and reusable schema. It can be decided quickly and comprehensibly, also at a later time, whether and how defects to be corrected are transferred back to development.}(S3). S2 confirmed that the differentiation enables to create a transparent and comprehensive \textit{``}[...] \textit{documentation that you can pull out even after a year or later }[...]\textit{."} Hence, the documentation, including detected defects which may have been graced for economical reasons, can be used to set up actions for latent faults \textit{``}[...]\textit{at an early stage or at least you can be aware that something can go wrong, without it having to go wrong first."}(S2).\\


\textbf{Effectiveness:} In terms of effectiveness, all three stakeholders of the postmortem analysis confirmed that it is possible to differentiate between the handling of fault detection and fault reaction from the handling of defect and its defect correction options. S2 stated that: \textit{``With the classification one can achieve a high accuracy in differentiation."} Moreover, S1 describes: \textit{``The classification enables us to distinguish more precisely between ``quick fix" and sustainable correction." }

All three stakeholders conclude, that the current organizational framework of the postmortem analysis is not yet an environment that is sufficiently conducive to learning. This illustrates the following statement (S1): \textit{``I think it would have been much easier if the organizational frame would have been different. The questioning of 5xWhy, was just limited compatible to it.} [...] \textit{It has disrupted the regular procedure of the postmortem analysis. Which is why it was not as efficient and took longer. But in terms of content it was definitely valuable."} This statement illustrates that the substantive goal was achieved. Concurrently the interviewees referred to aspects of time consumption and degree of difficulty, which brings us to the next evaluation criteria \textit{efficiency.}\\

\textbf{Efficiency:} The interviewees described that the application of the classification across different stakeholders and organizations was challenging and complicated. \textit{``You have individual views and a global view, which is presented simplified in retrospective. But it is not that easy in the analysis process... in reality it is nested with different responsibilities"} (S2).
In addition, S2 noted that at the beginning of an analysis appropriate data about anomalies at different system levels is not immediately available. Data can only be transferred into a form with sufficient information power through several iterations.

As described in paragraph \textit{effectiveness}, the efficiency was limited due to organizational frame. Therefore, we further discuss the relationship between the efficiency of applying the classification to the procedure of asking 5xWhy, that was mentioned by all three stakeholders.
S3 reflected that the application of the 5xWhy method goes beyond the boundaries between the effect chain and the tactical fault reactions during operation, without explicitly differentiating them. 
As a consequence: \textit{``With troubleshooting, outstanding unwanted effects ``faults" are reacted to with a fault reaction. When these unwanted effects are under control through a fault reaction, an interest in further actions related to sustainable correction decreases, even if the fault reaction has to be operated persistently."} (S3).

In terms of time consumption our classification performs worse than the previous postmortem. However, the application of our classification enabled us to identify faults with fault reactions and defects with options for sustainable correction. Therefore, defects related to product \footnote{For this work, a service is also a product.} (IoT system) quality
and defects related to capability of situation management could be identified.
From a strategic business point of view, the concepts of anomalies consequently allow to weigh up in present and also at a later stage in which activities are to be invested.

Further S3 highlighted that use of resources must be considered in relation to the effect achieved. \textit{``Symmetry and reuse of work is not the primary optimization goal of those who are involved in an postmortem analysis"}. This is where S3 identified a high potential for return on efficiency (see paragraph \textit{further Optimization}).

Throughout the interviews, we identified that in order to increase the time efficiency of the application further practical executions are needed.
We assume that the effort of applying the classification decreases with increasing usage and therefore in turn efficiency can be further increased.\\

\textbf{Satisfaction:}
All of the three stakeholders of the postmortem analysis stated to have a positive attitude towards the logic of the classification. S1 describes to feel secure in applying our concepts of anomalies and moreover to further distribute it across his team: \textit{``I like it a lot. That's why I use the concepts and terminology in our team. Also to get people used to it, because I think it´s value adding."} S3 commented the question of satisfaction by relating it to cultural aspects of performing postmortem analysis: \textit{``What I particularly like, is the fact that the classification refers to anomalies in the IoT system and not to people who are to blame for something."} S1 and S3 reflected on feeling confident in the application of the concepts. Uncertainty arose when participants noticed that other colleagues either had not understood the classification or worse, when the colleagues refused to differentiate due to mental effort, or in the course of transparency that was not intended. S2 added that: \textit{``}[...] \textit{indeed, one can hardly resist against the logic of the classification"}.\\

\textbf{Further Optimization:} We identified that the classification can be a foundation for an analysis that is not focused on blaming. However, it is not the solution to the cultural problem (S1): \textit{``With the concept you can also look for ``culprits" just as well as with the postmortem analysis before. At the end of the day it is crucial, what is done with the result of the analysis.}" S2 also discussed this point and stated: \textit{"If you disassemble everything in detail and you take a close look at what went wrong and where a defect is located, that is what people do not really enjoy. At this point you have to be especially careful not to be destructive." }

 S3 reflected how the execution of the 5xWhy in combination with the classification could be improved in terms of efficiency (S3): \textit{``Once the defect is detected or reported, so as assigned to a subset of components in arrangement, asking ``why" can be stopped. Which option for correction is finally executed, can be strategically planned outside of the postmortem analysis meetings, in the efficiency paradigm. Currently, in the postmortem meetings the options for correction are ``quickly found"... regardless of whether these corrections find a higher strategic use."}

It turned out that further improvement on training the classification, including practical instructions on how to apply it on a real-world customer problem are necessary. At this point it has to be ensured that people are not \textit{``}[...] \textit{overwhelmed with the differentiation"} (S2).

\subsection{Discussion - Stage 2}
The answers of the participants indicate, if the adaptation of the classification for anomalies is established and sufficiently practiced, it has the potential to be not only effective but also efficient. Misunderstandings in communications across different organisational units or silos can be drastically reduced. In addition, the reusability of the results of the classification fosters a continuous learning culture for iterative and inductive product development and operation.

Furthermore, the application of our concepts of anomalies is still challenging in varying degrees to the different stakeholders. \textit{``The colleagues with contact to the customer identified a benefit for themselves and quickly completed communication and thinking with the concepts that were new to them."} (S3). We assume that this is the case because the interest group in contact to the customer has the highest intention to create an overall view to control their value proposition. Further they have a high interest that defects are sustainably corrected to avoid further failures in their value provision to the customer. For teams at infrastructure level, with a more complex and dynamic view of the system, it is difficult to relate a fault and the related defect at the infrastructure level with a failures at the application level as described in an interview study from 2019 on observability and monitoring of distributed systems~\cite{niedermaier2019observability}.

We conclude that there is a need to focus on the organizational frame and on training of our classification for anomalies. The improvement of efficiency includes practical instructions in how to apply the classification for anomalies and how to improve its integration in the company specific procedure of the postmortem analysis. 

\section{\uppercase{Threats to Validity}}

Since we performed interviews for evaluating the quality in use of our classification, we anticipate some personal bias in the answers from the interviewees as threat to \textbf{internal validity}. To minimize this threat, we triangulated the answers of one interview of each stakeholder group. Further, the results of the classification have been classified and confirmed by the different stakeholders. We also compared the answers given in the interview with the observation of the classification meetings. Therefore, we consider the threat to internal validity has been further reduced.

Concerning \textbf{external validity}, we identified a risk that the classification and the results are specific to the case company context of Bosch. Hence, as we documented our adaptation on the IEEE Std. 1044-2009 for the system level as well as described the case context, we assume that our results can be generalized to similar IoT contexts outside of this specific case. We could further decrease this threat by applying it to other postmortem analysis and compare the results.

Since we have described the procedure of data collection and analysis, we consider that the study can be reproduced and a threat to \textbf{reliability} has been reduced. However, specific details about the incident report are sensitive and cannot be provided.

\section{\uppercase{Conclusion and Future Work}}
Building and operating IoT systems in open system context needs iterative and inductive learning to improve products. This is enabled by excluding unwanted effects in dynamics of customer context of use by automated fault reaction or defect correction. To collaborate along different product teams and organizations an ontology providing concepts of anomalies is needed.
With this work, we proposed and applied an adaptation of the IEEE Std. 1044-2009 for classifying system anomalies according to three architectural views: functional, dynamic and system structure view. Our adaptation allows to differentiate between the handling of fault detection and fault reaction in the context of situation management from the handling of defects and their defect correction options in the context of defect management. 
We have evaluated the quality in use with the quality sub-characteristics of effectiveness, efficiency and satisfaction of our adaptation by applying it to a postmortem analysis at Robert Bosch GmbH.

In terms of \textbf{effectiveness} we were able to differentiate between the concepts of failure and fault. This enabled us to identify missed fault reactions for controlling the system to keep the customer's trust during situation management and lets us improve fault tolerance and capability for situation management in future.
Moreover, we were able to differentiate between the concepts of fault and defect. This enabled us a differentiation between the actions for tactical fault fix from actions for sustainable defect correction.
In addition, the results of the postmortem analysis with the classification for anomalies are available in a reusable scheme and allow us to decide whether and how defects have to be corrected.

The \textbf{efficiency} of applying our classification was limited. The stakeholders described the application of the classification across different organizations as challenging, also due to the organizational frame. However, all interviewees highlighted the valuable content-related contribution. We anticipate that the effort and time of applying the classification will decrease with increasing usage to further improve efficiency.

The overall \textbf{satisfaction} of applying the classification was positive. The stakeholders stated a positive attitude towards the constructive usage of the concepts, in order to foster continuous learning. The classification can be foundation but will not be solution to difficulties in terms of silo mentality.

We identified that in industry there is a focus on the functional system view on the ``happy path". A dynamic system view, to control temporal effects, which enables fault detection and fault reaction is often missing. To be able to implement tactical options for fault reaction, we identified the need to enhance a system modelling. 

In order to sufficiently react to anomalies and keep the system in a controlled state performing operations, we have to observe and quantify the dynamic behaviour to indicate and detect faults. 
Therefore, we work on methods for observing anomalies to indicate and detect faults along the component-effect chains of a service throughout different organizational units,  performing development and operations. This should allow us to improve fault tolerance of complex IoT systems.
We think practitioners as well as researchers can take these concepts into account in the development of collaborative concepts in handling system anomalies along different organizational units.

\bibliographystyle{IEEEtran}
\bibliography{example.bib}

\end{document}